\documentclass{ptapap}
\usepackage{hyperref}
\def \ptaJournalDate{to be published}
\def \ptaJournalVolume{}
\def \ptaArticleURL{}
\def \ptaJournalName{PTA Proceedings}
\def \ArticleFirstPage{1}
\author{Klaudia Kowalczyk}[CAMK]
\author{Ewa L. {\L}okas}[CAMK]
\affil[CAMK]{Nicolaus Copernicus Astronomical Center\\Bartycka 18, 00--716 Warszawa, Poland}

\title{Estimating masses of dwarf spheroidal galaxies}

\begin{document}

\maketitle

\begin{abstract}
Precise measurements of mass in dark matter dominated dwarf spheroidal galaxies are of great importance for
testing the theories of structure formation. We use $N$-body simulations of the tidal evolution of a dwarf
galaxy orbiting the Milky Way to generate mock kinematical data sets and use them to test the reliability of
a simple mass estimator proposed by \citeauthor{wolf} The evolution of the initially disky dwarf galaxy
embedded in a dark matter halo was traced for 10 Gyr on a rather tight orbit. After about half of the time
a dwarf spheroidal galaxy is formed that retains some remnant rotation and a non-spherical shape.
Observing the triaxial galaxy along each of its principal axes we measure its half-light radius and
the line-of-sight velocity dispersion and use them to estimate the mass. We find that the mass is
significantly overestimated when the dwarf is seen along the longest axis of the stellar component and
underestimated when observed along the shortest axis. We provide a formula that quantifies the systematic
error in the estimated mass with respect to the true one as a function of the galaxy shape and line of sight.
\end{abstract}

\section{Introduction}
Determining accurate masses of dwarf spheroidal (dSph) galaxies usually requires strong assumptions about the
internal structure of galaxies, in particular about the anisotropy of stellar orbits, due to the degeneracies
between model parameters. Numerous methods including solving higher-order Jeans equations \citep{lokas2002}
or orbit superposition (\citealt{chaname}; \citealt{braddelsSculptor}) are being adopted to lift those
degeneracies. However, such approaches are significantly object-dependent and require significant amounts of
computation time so looking for simple methods is still appealing. One such method was recently proposed by
\citet{wolf} who offered a simple mass estimator based on observables such as the half-light radius an the
line-of-sight velocity dispersion. The mass in this case is estimated at a radius where the result is least
sensitive to the anisotropy.

In this work we test the reliability of this mass estimator in the context of the tidal stirring scenario for
the formation of dSph galaxies in the Local Group. In this scenario, dSphs form as a result of tidal
interaction of initially disky dwarfs with a bigger host, such as the Milky Way. Inherent in the process is
the formation of a triaxial stellar component that gradually becomes more spherical. We use an $N$-body
simulation following such an evolution to generate mock kinematic data sets and measure the observables
needed for the mass estimates. We demonstrate that the inherent non-sphericity of the objects introduces
systematic errors in their mass estimates.

\section{The simulation}
\label{The simulation}

Our simulation setup was composed of two galaxies: the Milky Way-like host and the dwarf galaxy, generated by
the procedures described in \citet{widrow} and \citet{widrowBlueprints}. Each object was modeled
using two components: an exponential disk and a cuspy NFW \citep{NFW} dark matter halo. To make the total
masses finite, the haloes were smoothly truncated at the radii close to the virial radii for both galaxies.
In total the simulation contained $8\times 10^5$ particles, $2\times 10^5$ particles per component for each
galaxy, with the numerical gravitational softening scales of $\epsilon_d=0.02$ kpc and $\epsilon_h=0.06$ kpc
for the dwarf's disk and halo and $\epsilon_D=0.05$ kpc and $\epsilon_H=2$ kpc for the Milky Way,
respectively.

For the dwarf galaxy we applied a model consistent with those used previously in similar studies
(\citealt{kazantzidisProgenitor}; \citealt{lokasTracks}). The dark matter halo had a mass $M_h=10^9\
\mathrm{M}_{\odot}$ and a concentration $c=20$. The disk had a mass $M_d=2\times 10^7\ \mathrm{M}_{\odot}$,
an exponential scale-length $R_d=0.41$ kpc and a thickness $z_d=0.2\ R_d$.

The host galaxy was based on
the model MWb of \citet{widrow}. Its dark matter halo had a mass $M_H=7.7\times 10^{11}\ \mathrm{M}_{\odot}$
and a concentration $c=27$ while its disk had a mass $3.4\times 10^{10}\ \mathrm{M}_{\odot}$, a length-scale
$R_D=2.82$ kpc and a thickness $z_D=0.44$ kpc. The neglect of the structural features of the Milky Way (like
the bar, bulge, thick/thin disk) was motivated by the simplicity and their relatively small impact on the
evolution of the dwarf which is of main interest here.

The dwarf galaxy was initially
placed at the apocenter of the eccentric, rather tight orbit with apocenter $r_{apo}=100$ kpc and pericenter
$r_{peri}=20$ kpc, which induced its faster (when compared to more extended orbits) transition towards the
spherical shape (\citealt{lokasShapes}; \citealt{kazantzidisShallowProfiles}). The evolution of the system
was followed for $10$ Gyr with the GADGET-2 $N$-body code (\citealt{springelNA}; \citealt{springelCode}) and
the outputs were saved every $0.05$ Gyr, giving 201 outputs in total.

\begin{figure}
\begin{center}
 \includegraphics[width=\textwidth, trim=40 490 20 25, clip]{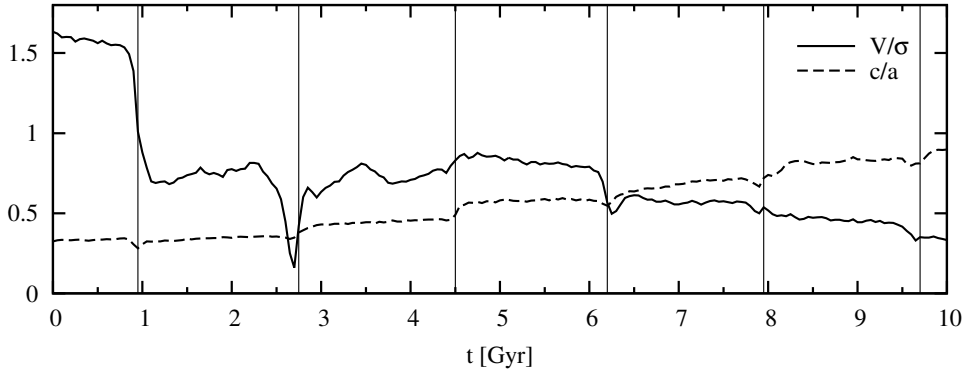}
\caption{The evolution of the amount of rotation with respect to random motions, $V/\sigma$ (solid line), and
the ratio of the shortest to the longest axis $c/a$ (dashed line) of the stellar component of the dwarf galaxy
in time. Both quantities were measured within a fixed radius $r<0.5$ kpc. Thin vertical lines indicate
pericenter passages.}
\label{fig:param_10gyr}
\end{center}
\end{figure}

Figure \ref{fig:param_10gyr} shows
the evolution of the stellar kinematics and the shape of the stellar component of the dwarf galaxy measured
within a fixed radius $r<0.5$ kpc therefore probing its central region. The quantity $V/\sigma$ plotted with a
solid line is the ratio of the mean rotation velocity around the shortest axis to the 1D velocity dispersion
(calculated by averaging the dispersions measured along three spherical coordinates) so it quantifies the
amount of the ordered versus random motion. The shape is presented with a dashed line in terms of the ratio
of the shortest to the longest axis of the stellar component $c/a$ determined using the inertia tensor.

\section{Estimating masses}
\label{Estimating masses}

We assume that a dSph galaxy forms when the amount of stellar rotation drops below half of its initial value
and the dwarf galaxy shape is sufficiently spherical, i.e. the shortest-to-longest axis ratio is $c/a>0.5$.
Adopting this criteria, we selected 110 simulation outputs (in the time range $t=4.55 - 10$ Gyr) out of the
total 201 outputs saved (see Figure \ref{fig:param_10gyr}). For each selected output we determined the
principal axes of the stellar component and the dwarf was rotated so that the $x$ axis was oriented along the
major, the $y$ axis along the intermediate, and the $z$ axis along the shortest axis.

\begin{figure}
\begin{center}
 \includegraphics[width=\textwidth, trim=10 340 20 25, clip]{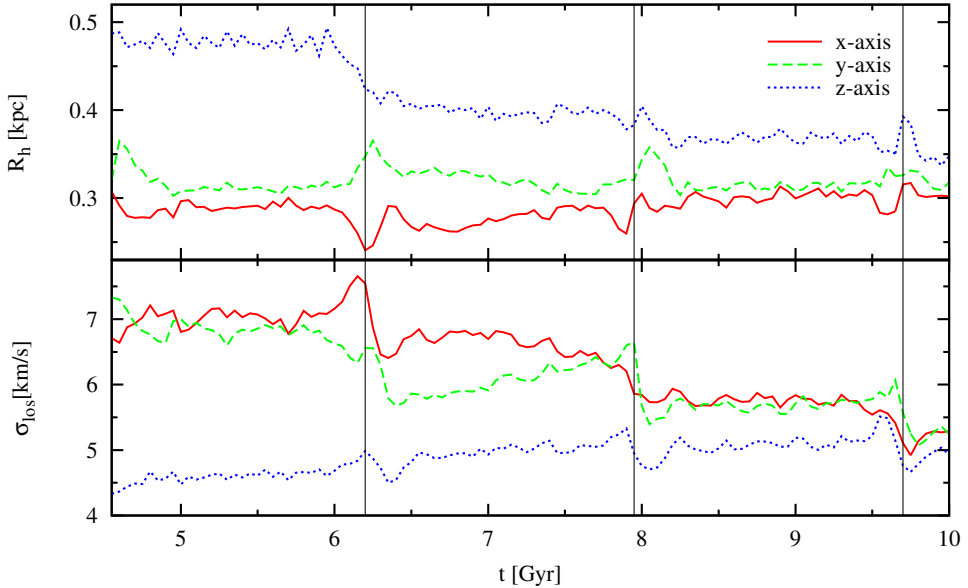}
\caption{The fitted values of the 2D projected half-light radii (top panel) and the line-of-sight velocity
dispersions measured within the half-light radii (bottom panel) as a function of time. Red solid lines, green
dashed lines and blue dotted lines show the results obtained for the observations along the longest ($x$),
the intermediate ($y$) and the shortest ($z$) axis of the stellar component, respectively. Thin vertical
lines indicate pericenter passages.}
\label{fig:fit_sig}
\end{center}
\end{figure}

Next, we `observed'
the dwarf along each of these axes creating for each line of sight a mock data set including the projected
stellar positions and radial velocities as would be available for a distant observer. The stellar positions
were binned equally in the logarithm of the projected radius $\log R$ to measure the number density profile
and to each such profile we fitted the projected Plummer distribution
\begin{equation}
	\Sigma(R)=\frac{NR_h^2}{\pi(R^2+R_h^2)}
\end{equation}
by adjusting the projected half-light radius $R_h$ and the normalization $N$.
We then measured the line-of-sight velocity dispersion $\sigma_{los}$ within $R_h$ removing for each sample
the stars stripped earlier by the tidal force with a converging $3\sigma$ clipping procedure until no more
stars were removed. The estimated values of $R_h$ and $\sigma_{los}$ for all selected outputs and
different lines of sight are plotted as a function of time in the top and bottom panels of Figure
\ref{fig:fit_sig}, respectively.

Treating each output as an individual dwarf galaxy, we used the measurements to estimate the masses
of the dwarfs by applying the formula proposed by \citet{wolf}:
\begin{equation}
	M_{est}(r_3)=3 \ G^{-1} \sigma_{los}^2 r_3 = 3.7\ G^{-1} \sigma_{los}^2 R_h
\end{equation}
where $r_3$ is the 3D radius found to contain mass least dependent on the orbital anisotropy. The radius $r_3$
is related to the 3D half-light radius $r_h$ and to the 2D projected half-light radius $R_h$ for the adopted
Plummer profile by $r_3/r_h=0.94$ and $r_h/R_h=1.305$, so that $r_3=fR_h$ where $f=1.23$ (see the Appendix in
\citealt{wolf}). We then compared the masses estimated in this way with the real masses ($M_{true}$)
contained within $fR_h$ of the simulated dwarfs.

\begin{figure}
\begin{center}
 \includegraphics[width=\textwidth, trim=20 190 20 25, clip]{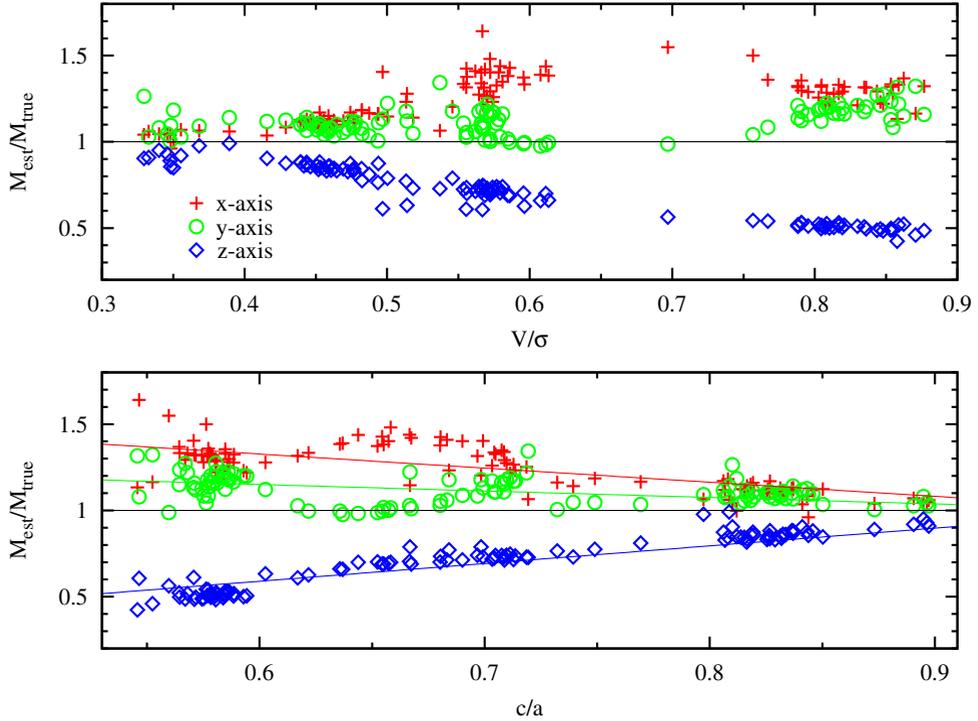}
\caption{The ratio of the estimated to the true masses as a function of the amount of rotation $V/\sigma$
(top panel) and the shape in terms of the ratio of the shortest to the longest axis $c/a$ (bottom panel).
Symbols of different shape and color (red crosses, green circles and blue squares) present the results
obtained for the observations along the longest ($x$), the intermediate ($y$) and the shortest ($z$) axis of
the stellar component, respectively. The horizontal black line indicates $M_{est}/M_{true}=1$ (no bias). Thin
lines of different color show the best-fitting linear relations between $M_{est}/M_{true}$ and $c/a$.}
\label{fig:ratio}
\end{center}
\end{figure}

We calculated the mean values and dispersions of the
ratio $M_{est}/M_{true}$ averaged over the whole sample for each line of sight separately obtaining: $1.24\pm
0.02$ for $x$ axis, $1.12\pm 0.01$ for $y$ axis and $0.70\pm 0.02$ for $z$ axis. In Figure \ref{fig:ratio} we
show the ratio $M_{est}/M_{true}$ for each line of sight as a function of the two global properties of the
dwarfs to which we trace the correlation: the amount of rotation $V/\sigma$ (top panel) and the shape of the
stellar component in terms of the ratio of the shortest to the longest axis $c/a$ (bottom panel). With
thin lines of different color we show the best-fitting linear relations of the form
\begin{equation}
	M_{est}/M_{true}(c/a) = \alpha  (c/a) + \beta
\end{equation}
with a constraint $M_{est}/M_{true}(c/a=1)=1$. The
obtained values of the slopes are: $\alpha=-0.82$ for the $x$ axis, $\alpha=-0.38$ for the $y$ axis and
$\alpha=1.03$ for the $z$ axis, respectively. Clearly, the estimated masses are less biased for more
spherical, non-rotating galaxies.

\section{Discussion}
\label{Discussion}
Using an $N$-body simulation of the tidal evolution of a dwarf galaxy placed on an eccentric orbit around a
Milky Way-like host we studied systematic errors in estimates of mass of dSph galaxies contained within a
radius of the order of a half-light radius. Due to a large
number of stars in the samples, with positions and velocities known to arbitrary accuracy of the numerical
calculations, the statistical errors are diminished, revealing uncertainties underlying the method itself.

We have demonstrated the impact of the triaxiality of the
stellar component on the obtained results. We have shown that the estimated masses can be either over- or
underestimated depending on the line of sight along which the dwarf is observed. Similarly to the results
presented in \citet{kowalczykSlopes} our estimated mass values are systematically larger than the actual ones
for the observations performed along the longest axis of the stellar component and smaller for the
observations along the shortest axis. The best agreement is achieved when we observe the dwarf along its
intermediate axis.

In this work we used a new simulation of tidal stirring with the dwarf galaxy placed on a tighter orbit than
in \citet{kowalczykSlopes}. As a result, a dSph galaxy was formed before half of the total simulation time and
became almost spherical at the end of the evolution. This allowed us to study dwarf realizations with a wider
range of key parameters such as the shape $c/a$ and the amount of remnant rotation $V/\sigma$. In addition,
after the formation time ($t=4.55$ Gyr) these quantities turned out to depend almost monotonically on time
($c/a$ increasing and $V/\sigma$ decreasing) giving us the opportunity to notice systematic dependence of the
results of mass estimates on these quantities.
In particular, we now notice a very clear dependence of the accuracy of mass estimates on the shape parameter
$c/a$ with an almost linear relation between $M_{est}/M_{true}$ and $c/a$, and the convergence of the
estimated masses to true ones when the object becomes spherical. The dependence on the amount of rotation is
similar although less obvious and we believe it is to some extent a derivative of the dependence on the
shape or at least the two are strongly correlated. It has been shown here and e.g. by
\citet{klimentowskiDSphFormation} that the change of shape towards a more spherical one is always accompanied
by the loss of rotation.

\section*{Acknowledgements}
This research was supported in part by PL-Grid Infrastructure, the Polish National 
Science Centre under grant 2013/10/A/ST9/00023 and the Polish Ministry of Science and Higher Education under grant
0149/DIA/2013/42 within the Diamond Grant Programme for years 2013-2017. We thank L. Widrow for providing
procedures to generate $N$-body models of galaxies for initial conditions.

\bibliographystyle{ptapap}
\bibliography{references}

\end{document}